\documentclass[aps,pre,preprint,floatfix]{revtex4-1}

\usepackage{amsmath}
\usepackage{mathrsfs}
\usepackage{amssymb}
\usepackage{graphicx}
\usepackage{dcolumn}
\usepackage{bm}
\usepackage[usenames,dvipsnames]{xcolor}
\usepackage{multirow}
\usepackage{hyperref}
\usepackage{epstopdf}

\usepackage[utf8]{inputenc}
\usepackage[T1]{fontenc}
\usepackage{microtype}
\usepackage{txfonts}
\usepackage{ulem}
\usepackage{tabularx}

\usepackage{url,hyperref}
\hypersetup{colorlinks=true, linkcolor=blue, urlcolor=blue, citecolor=blue}

\def\brho{{\boldsymbol{\rho}}}

\usepackage{epstopdf}

\begin{document}

%-------------------------------------------------------------------------------

\title{Curvature effects in charge-regulated lipid bilayers}

\author{Petch Khunpetch$^{1}$}
\thanks{Equal contribution and shared co-first authorship}
\author{Arghya Majee$^{2,3}$}
\thanks{Equal contribution and shared co-first authorship}
\author{Rudolf Podgornik$^{4,5,6}$}
\affiliation{$^1$School of Physical Sciences, University of Chinese Academy of Sciences, Beijing, China\\ 
             $^2$Max Planck Institute for Intelligent Systems, Stuttgart, Germany\\
             $^3$IV. Institute for Theoretical Physics, University of Stuttgart, Germany\\
             $^4$Kavli Institute for Theoretical Sciences, University of Chinese Academy of Sciences, Beijing, China  
             \& CAS Key Laboratory of Soft Matter Physics, Institute of Physics, Chinese Academy of Sciences, Beijing, China\\
             $^5$Wenzhou Institute of the University of Chinese Academy of Sciences, Wenzhou, Zhejiang, China\\
             $^6$Department of Theoretical Physics, Jo\v zef Stefan Institute, Ljubljana, Slovenia 
             \& Department of Physics, Faculty of Mathematics and Physics, University of Ljubljana, Ljubljana, Slovenia}

\date{February 16, 2022}

\begin{abstract}
We formulate a theory of electrostatic interactions in lipid 
bilayer membranes where both monolayer leaflets contain 
dissociable moieties that are subject to charge regulation. 
We specifically investigate the coupling between membrane 
curvature and charge regulation of a lipid bilayer vesicle 
using both the linear Debye-H\" uckel (DH) and the non-linear 
Poisson-Boltzmann (PB) theory. We find that charge regulation 
of an otherwise symmetric bilayer membrane can induce 
{\sl charge symmetry breaking, non-linear flexoelectricity} 
and {\sl anomalous curvature dependence of free energy}. 
The pH effects investigated go beyond the paradigm of 
electrostatic renormalization of the mechano-elastic 
properties of membranes.
\end{abstract}

\maketitle

%-------------------------------------------------------------------------------

\section{\label{sec:intro}Introduction}
The effects of electrostatic interactions \cite{Mar21,Bos21} on 
properties of lipid bilayer membranes are significant, ubiquitous 
and biologically \cite{Hon08} and biomedically \cite{Ewe21} 
relevant, since naturally occurring as well as technologically 
relevant lipids usually carry numerous dissociable or polar 
groups \cite{Cev90}, and lipid membrane charge density has 
been identified as a universal parameter that quantifies the 
transfection efficiency of  lamellar cationic lipid–DNA 
complexes \cite{Ewe21}. The particular relevance of electrostatic 
interactions in lipidic systems very clearly transpired from 
the structural principles of the two vaccines approved against 
COVID-19 in December 2020 and based on the cationic lipid 
vesicle vector and the anionic mRNA payload developed by 
Pfizer/BioNTech \cite{Mul20}  and Moderna \cite{Bad21}. 
In general, charged lipids capable of modulating their charge 
depending on the bathing environmental conditions, 
have been recognized as a key component of {\sl ionizable 
lipid nanoparticles} which are currently acknowledged as 
one of the most advanced non-viral vectors for the efficient 
delivery of nucleic acids \cite{Sch21}. 

While electrostatic interactions in bio-soft matter are 
universal \cite{Hol01}, there is a fundamental difference 
between the standard colloid electrostatics and membrane 
electrostatics \cite{Bor01}, in the sense that the membrane 
charge, just as the protein charge \cite{Lun05, Bar09, Bar17}, 
depends on the solution  environment of the lipid \cite{Avn18} 
and its changes can engender also changes in the shape of 
the lipids and concomitant structure of lipid assemblies. 
One example of this solution environment effect would be 
changes in protonation/deprotonation  equilibria of the 
dissociable phospholipid moieties depending on the solution 
pH, that in general affect the charge of lipids' headgroup, 
and another would be the modification of the strength of 
electrostatic interactions wrought by the ionic strength of 
the bathing aqueous solution that affects the ionic screening. 
In particular, for therapeutic gene delivery \cite{Pod08} 
ionizable lipids are designed to have positive charges at 
acidic pH in the production stage to ensure a strong 
interaction with nucleic acids, after which they should 
become almost neutral under physiological conditions 
($\mathrm{pH}~7.4$) to prevent sequestration in the liver, and 
should finally turn positive again upon being introduced 
into a typical acidic environment of the endosomes, 
promoting the delivery of the genetic cargo, followed 
finally by release of the genetic cargo at neutral $\mathrm{pH}$ 
enabling the cargo to interact with the cell machinery \cite{Sch21}. 

To connect the dissociation equilibria of dissociable 
surface groups and electrostatic fields, Ninham and 
Parsegian (NP) \cite{Nin71} introduced the {\sl charge 
regulation (CR) mechanism}, that couples the Langmuir 
isotherm model of the local charge association/dissociation 
process with the local electrostatic potential \cite{Mar21}, 
resulting in an electrostatic self-consistent boundary 
condition. The NP model can be generalized to include 
other adsorption isotherms, depending on the detailed 
nature of the dissociation process \cite{Pod18}.

The protonation/deprotonation equilibria at the dissociable 
surface groups in phospholipid membranes  involve local $\mathrm{pH}$ 
and local bathing solution ion concentrations, which in 
general differ from the bulk  conditions \cite{Nap14}. 
This implies that the changes in the bathing solution 
properties will have a pronounced effect on the effective 
charge of the membrane, thus modifying $\mathrm{pH}$ 
sensing and $\mathrm{pH}$ response of lipid membranes \cite{Ang18}. 
In fact, $\mathrm{pH}$-dependence of the lipid charging state 
can directly enable an ionizable lipid with a dissociation 
constant $\mathrm{p}K_a \sim 6.5$ to be neutral in the 
blood circulation, thus preserving a bilayer structure, and 
then revert to its charged protonated form at an endo-lysosomal 
$\mathrm{pH}$, consequently reverting to an hexagonal 
phase upon contact with anionic membrane lipids \cite{Jay12}. 

\begin{figure*}[!t]
\centering
\includegraphics[width=\columnwidth]{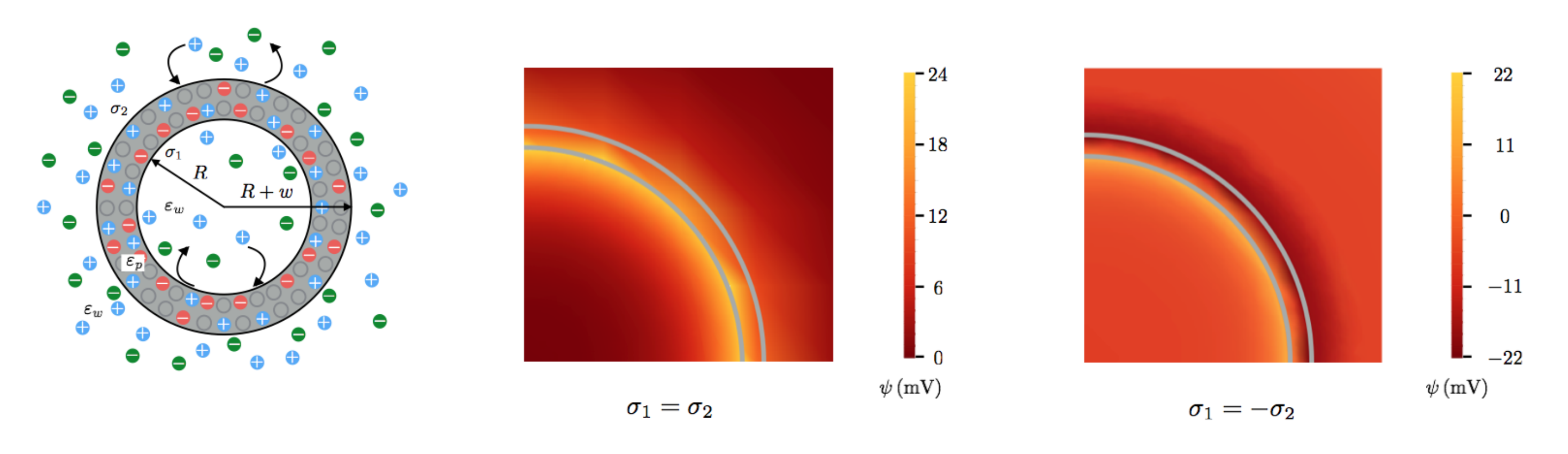}
  \caption{{Left: A schematic rendition of a spherical charged 
  vesicle of inner and outer radii $R$ and $R+w$, respectively, 
  immersed in a salt solution. The inner (outer) surfaces of 
  the bilayer carry ionizable groups and are characterized by a 
  charge density $\sigma_1$ ($\sigma_2$). The lipid bilayer 
  (indicated in gray) of width $w$ and dielectric constant 
  $\varepsilon_p \simeq 5$ separates the interior of the shell 
  from the outer region, both of which are typically aqueous 
  phases characterized by a dielectric constant 
  $\varepsilon_w \simeq 80$. All the ionic species in both regions 
  are assumed to be in chemical equilibrium defined by the 
  bulk chemical potentials. Both the inner and outer surfaces 
  of the bilayer carry fixed numbers of negatively charged 
  surface groups (denoted in red) as well as neutral sites where 
  (de)protonation reactions (indicated by the arrows) can take 
  place leading to surface charge densities $\sigma_1$ and 
  $\sigma_2$. Center and right: the color-coded electrostatic 
  potential profile for the symmetric case, corresponding to 
  $\sigma_1=\sigma_2=0.5 ~{e/\rm nm}^2$ and for the asymmetric 
  case with $\sigma_1=-\sigma_2=0.5 ~{e/\rm nm}^2$ 
  obtained from the charge regulation calculation as explained 
  in the main text. For both cases, $R=20\,\rm{nm}$, $w=1\,\rm{nm}$, 
  and $\kappa_D=1.22\,\rm {nm}^{-1}$ are used.}}
  \label{Fig:1}
\end{figure*}

In what follows we will show that even in the case of a chemically 
symmetric curved membrane, {\sl i.e.} with the same lipid composition 
on the inner and outer leaflets, the CR process introduces a charge 
asymmetry and an out-of-plane membrane polarization vector in a 
highly non-linear fashion, characterized by a sudden symmetry 
breaking transition involving in all other respects {\sl chemically identical} 
outer and inner leaflet surfaces of the membrane. This phenomenology 
closely parallels the recently discovered symmetry breaking charge 
transitions in interactions between pairs \cite{Maj18} or stacks \cite{Maj19} 
of charged membranes as well as the complex coacervation driven by 
the charge symmetry broken states in solutions of macroions with 
dissociable surface moieties \cite{Maj20}.

{\sl Charged lipids.} Different naturally occurring charged lipid species, 
an important fraction of which is anionic \cite{Gal21}, represent a 
notable part of biological membranes, and consequently their role in 
the structural, dynamical and mechanical properties of model lipid 
bilayers as well as their effect on the membrane proteins is of great 
biological interest \cite{Poy16}. The less commonly occuring cationic 
lipids, on the other hand, have played an important role as the main 
charged component of the cationic liposome vectors for nucleic acid 
therapeutics \cite{Pod08}. 

The charge of phospholipid polar heads originates in deprotonated 
phosphate groups, protonated amine group, and deprotonated carboxylate 
group, with the corresponding dissociation constants expected to lie in 
the region of $0 \leq \mathrm{p}K_a \leq 2$ for the primary phosphate 
group, in the region $6 \leq \mathrm{p}K_a \leq 7$ for the secondary 
phosphate group, in the region $3 \leq \mathrm{p}K_a \leq 5$ for the 
carboxylate groups, and $9 \leq \mathrm{p}K_a \leq 11$ for the primary 
amine group \cite{Mar13}. Since in the subsequent analysis the effect 
of the electrostatic interactions - including the local polarity, the ionic 
strength and the distribution of neighboring dissociable groups - will 
be considered explicitly, we consider only the intrinsic $\mathrm{p}K_a$, 
noting that even in the absence of CR certain structural and continuum 
electrostatic effects are sometimes included in the calculation of the 
dissociation constants when determined by {\sl e.g.} PROPKA \cite{Ols11}.

Of the different phospholipids, phosphatidylcholines have very 
low $\mathrm{p}K_a$ and can be considered as undissociated, 
but in particular phosphatidylserine, phosphatidylinositol, as well 
as phosphatidylglycerol, phosphatidylethanolamine, cardiolipin 
and phosphatidic acid all contain at least one dissociable moiety 
at relevant pH conditions and should be considered as potentially 
charged \cite{Cev18}. Among the cationic lipids DOTAP 
(1,2-dioleoyl-3-trimethylammonium-propane) remains charged 
irrespective of pH, while the amine groups of other cationic lipids 
acquire their charge by protonation only below a certain pH \cite{Lev90}. 
As an extreme case one could mention the custom-synthesized 
cationic lipid MVLBG2 with a headgroup that bears no less than 
16 positive charges at full protonation \cite{Ewe21}. In ionizable 
lipids the quaternary ammonium head of cationic lipids can be 
in addition substituted with a titratable molecular moiety with an 
engineered  dissociation constant that would ensure the charge 
would be mouldable by the environmental $\mathrm{pH}$ \cite{Sch21}. 
Among the ionizable lipids that respond strongly to pH in the relevant 
range one can list DODAP (1,2-dioleoyl-3-dimethylammonium propane), 
DLinDMA, DLin-KC2-DMA and finally DLin-MC3-DMA, all of them 
used for proper lipid vector formation in cytosolic delivery of 
therapeutic cargo as reviewed recently in Ref. \cite{Sch21}.  

{\sl Protonation and pH effects.} Detection and response of the cells 
to either global pH environment or local pH gradients is crucial for 
optimal cellular metabolism, growth and proliferation \cite{Cas09}. 
The sensing of pH heterogeneities is important for initiating  migration, 
polarization  and deformations of lipid bilayer assemblies \cite{Ang18} 
and can be driven by protonation changes of phospholipids that can 
furthermore directly affect the membrane curvature \cite{Ben12}. In 
fact pH regulation of phospholipid charges have been implicated in 
pH-induced migration, pH-induced bilayer local or global deformation, 
as well as pH-induced vesicle chemical polarization \cite{Ang18}. 
The fact that the charge state of many phospholipids depends on the 
solution pH implicates by and large a charge regulation mechanism 
as indeed first proposed for surfaces containing dissociable groups 
by Ninham and Parsegian \cite{Nin71}. 

In addition, the phospholipid protonation state and the membrane 
charge distribution can modify the interaction energy with a protein 
on approach to the membrane \cite{Lun13}. This effect too is due 
to perturbed protonation equilibrium at the membrane titratable sites, 
with phospholipid charge regulation determining the strength and 
even the sign of the protein-membrane interaction in variable chemical 
environments \cite{Jav21}. Charge regulation can thus be seen as a 
{\sl biological $\mathrm{pH}$-sensitive switch} for protein binding 
to phospholipid membranes and  thus conferring signaling functions 
to different types of lipids \cite{Tan18}.

{\sl Curvature and lipid dissociation effects.} Global and local changes 
in pH, inducing changes in the charged states of lipids, can affect lipid 
vesicle shape deformations as well as phase-separated domain 
formation \cite{Ang18}. Usually these effects are framed within the 
electrostatic contribution to the mechano-elastic properties of membranes, 
such as surface tension and bending rigidity,  which have been reviewed 
in detail, see e.g. Refs. \cite{And95}, \cite{Fog99} and \cite{Gal21}. 
Electrostatic coupling of membrane charges leads to a salt concentration 
dependent increase in the bending rigidity of the charged membrane 
compared to a charge-free membrane \cite{Dim19}. Usually the details 
of lipid dissociation mechanism are not considered in these works and 
analyses going beyond the fixed charge assumption, incorporating 
direct lipid charge dissociation models, are less common but have 
been nevertheless invoked when modelling the flexoelectric properties 
of membranes \cite{Pet99}. In fact Langmuir adsorption isotherm was 
used to estimate the change in the surface density of counterions on 
the inner and outer leaflet surface of the phospholipid membrane 
induced by the coupling between curvature and electrostatics \cite{Hri91}. 

In what follows we will present a detailed analysis of the charge 
regulation effects in a system, see Fig. \ref{Fig:1}, comprised of 
a single spherical unilamellar lipid vesicle of a fixed phospholipid 
membrane thickness $w\simeq 4\,\mathrm{nm}$, dielectric constant 
$\varepsilon_p \simeq 5$ and a variable inner radius $R$, 
immersed in a simple univalent aqueous electrolyte solution, of 
dielectric constant $\varepsilon_w \simeq 80$. We will use the full 
Poisson-Boltzmann (PB) theory to evaluate the electrostatic part 
of the free energy, as well as the often invoked and easier to 
implement linearized Debye-H\"uckel (DH) theory in the second 
order curvature expansion approximation. This will allow us to 
ascertain which effects are non-linear in nature as far as 
electrostatics is concerned. We will show that charge regulation 
of an otherwise symmetric bilayer membrane induces three 
important phenomena: {\sl charge symmetry breaking, non-linear 
flexoelectricity and anomalous curvature dependence of free energy} 
and that in general the pH effects go beyond the paradigm of 
electrostatic renormalization of the mechano-elastic properties 
of membranes.

The outline of the paper is as follows: In Sec.~\ref{sec:model_formalism}, 
we describe the details of the charge regulation model and formalism 
for this study. Section~\ref{sec:results} is devoted for the results of the 
study. Finally, the discussion of the results of the curvature effects in 
charge-regulated lipid bilayers is given in Sec.~\ref{sec:discussion}. 

\section{\label{sec:model_formalism}Model and formalism}

\subsection{\label{sec:model}Model}

The description of charge regulation is model specific and there 
is no universal model that would be applicable to any situation 
encountered in biophysical systems. We base our model on the 
preponderance of three different types of interactions that can 
be deemed as important in the context of ionizable lipids: 
(i) electrostatic interactions, (ii) non-electrostatic adsorption 
interactions and (iii) adsorption site entropy. 

We consider a charged spherical vesicle with a salt solution on the 
two sides of its bilayer membrane as shown in Fig.~\ref{Fig:1}. The 
inner and the outer radii of the charged shell are given by $R_1=R$ 
and $R_2=R+w$, respectively. While the model considers a globally 
curved vesicle, it nevertheless provides also some insight into the 
local curvature deformations which are formally much less accessible 
to detailed calculations. The connection is analogous to the global and 
local curvature effects in stiff polyelectrolytes such as DNA \cite{Bre82}.

The lipid bilayer is composed of two charge-regulated monolayers 
with charge densities $\sigma_{1,2}$. These charge densities stem from 
(de)protonation reactions or dissociation/adsorption of mobile charges 
taking place at chargeable sites present in each monolayer. We assume 
that $n_{1,2}$ is the number of negative lipid heads per surface area 
and $\Theta n_{1,2}$ is the number of neutral lipid heads per surface 
area of the inner and outer monolayers, respectively. The phospholipid 
bilayers are assumed to be incompressible (see Ref. \cite{Dim19} and 
references therein) so that 
\begin{align*}
n_i = n_{0} \left( \frac{R_0}{R_i}\right)^2,
\end{align*}
where $R_0$ is the radius of the neutral plane, $R_i = R_0 \pm w/2$,
while $-$ and $+$ occur for $i=1$ and $2$, and $n_{0}$ is the dissociable 
lipid density at the neutral plane. To the lowest order in mean curvature 
and thickness of the bilayer $w$, this can be expanded to obtain 
$n_i\simeq n_{0} \left(1 \pm \frac{w}{R_0} + \dots\right)$.

Furthermore, $\eta_{1,2}$ are the fractions of the neutral lipid heads 
where the adsorption/desorption of cations can take place on the inner 
and outer monolayers, respectively. By definition, $\eta_{1,2} \in [0, 1]$. 
$e$ is the elementary charge ($e > 0$). Note that here we have modified 
the model discussed in~\cite{Maj18}. The connection between lipid density, 
charge density and adsorbed fraction is as follows:
\begin{align}
\sigma_i =  -n_i e + \Theta n_i e\eta_i.
\label{sigma}
\end{align}

In our model, $\sigma_{1,2}$ and $\eta_{1,2}$ are assumed to be 
uniform over the two constituting monolayers of the membrane, 
and we specifically consider the case $\Theta=2$ for simplicity. 
This assumption reduces Eq.~\eqref{sigma} to
\begin{align}
\sigma_i = 2 n_i e \left(\eta_i - {\textstyle\frac{1}{2}}\right),
\label{sigma12}
\end{align}
with $- e n_{1,2} \leq \sigma_{1,2} \leq e n_{1,2}$, implying that the 
charge of the membrane can change sign as a consequence of a 
mixture and different levels of dissociation of aionic and cationic lipid 
components as well as membrane embedded proteins, in general. 
Other choices of $\Theta$ are of course also possible depending 
on the exact composition of the bilayer and the pertaining dissociation 
constants. This does not, however, impact the qualitative features 
of our results. The Debye screening length $(\lambda_D)$ varies 
from about $0.34\,\mathrm{nm}$ to $10.75\,\mathrm{nm}$ 
corresponding to the monovalent salt concentration ranging from 
$1-0.001\, \mathrm{M}$ \cite{Sho16}.

\begin{table}[t]
\small
  \caption{Different variables and their meanings. $\alpha$ and 
  $\chi$ are dimensionless energies expressed in the units of the 
  thermal energy and are of non-electrostatic origins.}
  \label{Table:1}
  \def\arraystretch{1.5}
  \begin{tabular*}{\textwidth}{@{\extracolsep{\fill}}ll}
    \hline
    \hline
    Symbol & Definition \\
    \hline
    $n_{1,2}$ & density of lipid headgroups\\
    $\eta_{1,2}$ & fractions of charged lipids; $0\leq\eta_{1,2}\leq 1$ \\
    $e$ & elementary charge $(e>0)$\\
    $\sigma_{1,2}$ & charge densities of the two monolayers; $\sigma_i=2e n_i \left(\eta_i-\frac{1}{2}\right)$ \\
    $R_{1,2}$ & radius of the monolayers; $R_{1,2}=R, R+w$ \\
    $w$ & width of the bilayer; $w=R_2-R_1$ \\
    $\alpha$ & (de)protonation energy; $\alpha=\left(\mathrm{p}K_a-\mathrm{pH}\right)\ln10$ \\
    $\chi$ & lateral pair interaction energy of occupied neighboring sites \\
    $\varepsilon_{w,p}$ & dielectric constants of water or lipid \\
    $\beta$ & inverse thermal energy; $\beta=1/\left(k_BT\right)$ \\
    $\lambda_D$ & Debye screening length; $\lambda_D^2=\varepsilon_w\varepsilon_0/\left(2n_I\beta e^2\right)$ \\
    $\kappa_D$ & inverse Debye length; $\kappa_D=1/\lambda_D$ \\
    $\ell_B$ & Bjerrum length; $\ell_B=\beta e^2/\left(4\pi\varepsilon_w\varepsilon_0\right)$ \\
    $h$ & dimensionless curvature; $h=1/\left(\kappa_DR\right)$ \\
    $n_I$ & bulk ionic strength \\
    $\psi$ & mean-field electrostatic potential\\
    \hline
    \hline
  \end{tabular*}
\end{table}

Charge regulation can be quantified either through the chemical 
equilibrium of dissociable sites at the  bounding surface or through 
the surface free energy if the interactions between the solution 
charges and the surface is characterized by short-range ion-specific 
interactions \cite{Pod18}. The latter seems more appropriate within 
the context of mean-field electrostatics.

In general, charge regulation is related to any non-trivial, {\sl i.e.} 
non-zero, form of the surface free energy describing different 
models of surface-ion solution interactions \cite{Bor01}. The single 
site dissociation model can be related to van’t Hoff adsorption 
isotherm, Langmuir (Henderson-Hasselbalch) adsorption 
isotherm, Frumkin-Fowler-Guggenheim adsorption isotherm 
and others \cite{Pod18}. Following the analysis of charged 
surfactant systems \cite{Har06} we base our phospholipid 
charge regulation model on the Frumkin-Fowler-Guggenheim 
isotherm \cite{Koo20} defined with the phenomenological free 
energy of adsorption sites at surface density $n$ in the units 
of thermal energy $k_BT=1/\beta$ as
\begin{align}
  \beta{\cal F}_{CR}\left[\eta\right] = n \oint\limits_{A} dA \bigg[- \alpha\eta({\brho}) -\frac{1}{2}{\chi} \eta({\brho})^2 + \eta({\brho}) \ln(\eta({\brho}))
+ (1-\eta({\brho}))\ln(1-\eta({\brho}))\bigg],
   \label{eq:1}
\end{align}
where the integral is over the surface with dissociable groups 
and ${\brho}$ is the radius vector on this surface. For a spherical 
surface of radius $R$, $|{\brho}|=R$. The dimensionless parameters 
$\alpha$ and $\chi$, both expressed in thermal units in the above 
definition, are phenomenological and describe the non-electrostatic 
parts of the adsorption energy and of the surface lateral Flory 
interaction strength, respectively. A more general, non-local version 
of the latter would contribute a term, 
\begin{align}
\oint_{A} dA ~\frac{1}{2}\chi\eta(\rho)^2 \longrightarrow \oint_{A} dA dA'~\frac{1}{2} \chi(\rho - \rho') \eta(\rho)\eta(\rho'),
\end{align}
to the free energy, which, however, we will not pursue in what follows, 
limiting ourselves to the simplified Frumkin-Fowler-Guggenheim isotherm 
of Eq.~\eqref{eq:1}. In polymer solution theory and regular solution theory 
$\chi$ is referred to as the Flory–Huggins interaction parameter while in 
the context of the Frumkin-Fowler-Guggenheim isotherm it is referred 
to as the Flory lateral interaction strength \cite{Koo20}.

The dependence of the adsorption energy $\alpha$ on the bulk 
concentration of the dissociation product (e.g., $p = $  protons, ions) 
is model specific \cite{Avn18}, but can be written as a sum of
\begin{align}
    \alpha &= \Delta g + \mu = \Delta g + k_BT \ln{a_{p}} 
    \label{alphadef}
\end{align}
where $\Delta g$ is the dissociation free energy difference, while the 
chemical potential $\mu = k_BT \ln{a_p}$ of the dissociation product 
is expressed through the absolute activity, $a_p$, that contains also 
the excess part dependent on the contribution of electrostatic interactions, 
see Ref.~\cite{Lan20}. On the mean-field PB level the chemical potential 
is given by the ideal gas form, which does not take into account the 
electrostatic interactions, with the activity proportional to the concentration. 
Therefore in this case 
\begin{align}
    \alpha &= k_BT~\ln{K_a} + k_BT \ln{[c^{+}]} 
    \label{alphadef11}
\end{align}
with the equilibrium constant $K_a$, defined standardly as 
$\Delta g = k_BT~\ln{K_a}$. For protonation/deprotonation reactions 
corresponding to $\mathrm{AH}^+ \rightleftharpoons \mathrm{A} + \mathrm{H}^+$ 
and $\mathrm{A}^- + \mathrm{H}^+ \rightleftharpoons \mathrm{AH}$ 
we then have $\mu = k_BT \ln{[\mathrm{H}^{+}]}$ and consequently 
\begin{align}
    \alpha &= \left(\mathrm{p}K_a-\mathrm{pH}\right)\ln 10,
    \label{alphadef1}
\end{align}
where now $\mathrm{pH}=-\log_{10}{[\mathrm{H}^{+}]}$. In what follows 
we will use $\alpha$ to quantify the non-electrostatic surface interactions 
with the forms Eqs.~\eqref{alphadef11},\eqref{alphadef1} implied for the 
general ion dissociation and protonation/deprotonation, respectively.

Furthermore, $\chi$, as in the related lattice  regular solutions theories 
(e.g., the Flory-Huggins theory \cite{Ter02}) describes the  short-range 
interactions between nearest neighbor adsorption sites on the macroion 
surface \cite{Avn20}. $\chi\geq0$ represents the tendency of 
protonated/deprotonated lipid headgroups on the macroion surface 
adsorption sites to phase separate into domains. The microscopic source 
of this lipid demixing energy could be due to some mismatch of 
head-group–head-group interactions, such as hydrogen bonding between 
neutral lipids, water-structuring forces, or nonelectrostatic ion-mediated 
interactions between lipids across two apposed bilayers for small 
interlamellar separations, see also Ref.~\cite{Har06}.

Adding the electrostatic energy, as will be done in the next section, 
our model will therefore incorporate all three fundamental interactions: 
electrostatic interactions, non-electrostatic interactions between the 
lipids and the solution ions as well as non-electrostatic interactions 
between the adsorbed ions, and the adsorption site entropy.

\subsection{\label{sec:electrostatic}Electrostatic free energy}

We start with the standard Poisson-Boltzmann free energy, or the 
Debye-H\"uckel free energy in the linearized case, that depends 
on the charges and the curvature. There are various ways to write 
down the PB free energy \cite{Mar21} and we choose the field 
description, with the radially varying  electrostatic potential $\psi(r)$ 
as the only relevant variable. The total electrostatic free energy 
of our model system is then  
\begin{align}
\beta {\cal F}_{ES} = &-\int dV \left[  \frac{1}{2}\beta \varepsilon_w\varepsilon_0\left(\frac{d\psi(r)}{dr}\right)^2 + 2 n_I  \big( \cosh{\beta e \psi(r)}-1\big)\right]\notag\\ 
&+ \beta \oint_{A_1} dA_1~\psi\left(R_1\right) \sigma_1 + \beta \oint_{A_2} dA_2~\psi\left(R_2\right) \sigma_2,
\label{electrostatic1}
\end{align}
where $n_I$ is the univalent electrolyte concentration in the bulk, 
$R_1 = R$ and $R_2 = R + w$, while the equilibrium value of $\psi(r)$ 
is obtained from the corresponding Euler-Lagrange equation. The 
volume integral extends over all the regions except the bilayer interior. 
The electrostatic part of the thermodynamic potential can then be 
written with the help of the {\sl Casimir charging formula} (for details 
see Ref.~\cite{Ver48}) which allows one to write the full thermodynamic 
potential of the system in the form of a surface integral 
\begin{align}
\beta {\cal F}_{ES}(\sigma_1, \sigma_2, R) =  \beta ~\sum\limits_{i=1}^2
4\pi R_i^2 \int\limits_0^{\sigma_i}d\sigma_i~\psi\left(R_i, \sigma_i\right),
\label{electrostatic1a}
\end{align}
where $i=1,2$ corresponds to inner/outer leaflets of the bilayer. In 
the above equation note the integral of the boundary potential over 
the surface charge densities, pertinent to the full non-linear PB theory. 
From here it also clearly 
follows that
\begin{align}
\frac{\partial \beta {\cal F}_{ES}(\sigma_1, \sigma_2)}{\partial \sigma_i} = \beta ~4\pi R_i^2~ \psi\left(R_i\right),
\label{BC}
\end{align}
which we will invoke in the mean-field form of the charge regulation 
boundary conditions to be introduced later.

A common approach to electrostatic effects in membranes is {\sl via} 
the DH approximation together with small curvature, second order 
expansion \cite{And95, Fog99, Gal21}. In the DH approximation, valid 
specifically for $\beta e \psi(r) \ll 1$, the corresponding expressions for 
the electrostatic free energy Eq.~\eqref{electrostatic1} simplifies 
considerably to 
\begin{align}
\beta {\cal F}_{ES} = -\frac{1}{2}\beta ~\varepsilon_w\epsilon_0 \int dV \Bigg(  \left(\frac{d\psi(r)}{dr}\right)^2 + \kappa_D^2  \psi(r)^2\Bigg) + \beta \oint_{A_1} dA_1~\psi\left(R_1\right) \sigma_1 + \beta \oint_{A_2} dA_2~\psi\left(R_2\right) \sigma_2,
\label{electrostatic2}
\end{align}
where the inverse square of the Debye screening length $\lambda_D$ is 
given by $$\kappa_D^{2} = 2 n_I \beta e^2/\varepsilon_w\varepsilon_0$$ 
and the volume integral again extends over all the regions except the 
bilayer interior. Eq.~\eqref{electrostatic2} is then further reduced to 
\begin{align}
\beta {\cal F}_{ES}(\sigma_1, \sigma_2, R)= \beta~ 2\pi\sum\limits_{i=1}^2 R_i^2\sigma_i\psi\left(R_i, \sigma_i\right),
\label{nkcajwgscfkj3}
\end{align}
since the potentials are linear functions of the charge density and the 
integrals over surface charge densities in the Casimir formula can be 
evaluated explicitly. Within the phospholipid core of the bilayer the 
electrostatic energy is simply
\begin{align}
\beta {\cal F}_{ES} =  - \frac{1}{2}\beta \varepsilon_p\varepsilon_0 \int dV  \left(\frac{d\psi(r)}{dr}\right)^2 
 + \beta \oint_{A_1} dA_1~\psi\left(R_1\right) \sigma_1 + \beta \oint_{A_2} dA_2~\psi\left(R_2\right) \sigma_2,
\label{electrostatic11}
\end{align}
where the volume integral now extends over the bilayer interior. While 
the free energies Eq.~\eqref{electrostatic1} and  Eq.~\eqref{electrostatic2} 
imply the PB and the DH equation in the regions accessible to electrolyte 
ions \cite{Maj16}, respectively, Eq.~\eqref{electrostatic11} leads to the 
standard Laplace equation inside the lipid dielectric core.

The explicit DH expression for a charged spherical dielectric shell 
electrostatic free energy as a function of the radius of curvature was 
obtained in an analytical form in Ref. \cite{Sib07}. This was expanded 
up to the inverse quadratic order in curvature in Eq. 23 of Ref.~\cite{Sho16}. 
This is the analytical formula that we use in the DH part of our calculations. 
The second order curvature expansion was standardly taken as a point 
of departure for the electrostatic renormalization of the mechanical 
properties of membranes, such as surface tension and bending rigidity 
\cite{Win88, Mit89, Lek90, Dup90, Har92}. The methodology of solving 
the non-linear PB theory is the same as used in our previous publications 
\cite{Maj16, Maj18, Maj19, Maj20} and will not be reiterated again.

\subsection{\label{sec:cr}Charge regulation free energy}

Assuming that the inner and outer membrane surfaces are chemically 
identical we presume that the surface charge regulation process can 
be described by the Frumkin-Fowler-Guggenheim adsorption isotherm, 
including the adsorption energy, the interaction energy between 
adsorbed species and the site entropy. 

The corresponding free energy is then given by Eq.~\eqref{eq:1} so 
that the charge regulation free energy density of the inner and outer 
membrane surfaces denoted by $i=1,2$ read as
\begin{align}
  \beta {{\cal F}_{CR}(n_{i}; \eta_{i})} = n_{i} 4\pi R_{i}^{2} \left[ - \alpha\eta_{i} -\frac{\chi}{2}\eta_{i}^2+\eta_{i}\ln\eta_{i}
  + (1-\eta_{i})\ln(1-\eta_{i})\right].
  \label{cr1}
\end{align}  
This can be furthermore normalized w.r.t. the inner area $4\pi R^{2}$ 
which is used later. Formally, the first two terms in the free energy are 
enthalpic in origin, while the other terms are the mixing entropy of 
charged sites with the surface area fraction $\eta$ and neutralized 
sites with the surface area fraction $1-\eta$. The phenomenological 
constants of course contain enthalpic as well as entropic contributions 
from other microscopic order parameters which are not considered 
explicitly.

As we already stated, the dissociation constants of anionic phospholipids 
such as PS, PE, or PA lie in the region of $0 \leq \mathrm{p}K_a \leq 11$ 
\cite{Mar13,Cev18}, while for cationic lipids such as DLin-KC2-DMA, 
DLin-MC3-DMA, DLin-DMA, DODMA, and DODAP the dissociation 
constants have been estimated to lie in the region of 
$5\leq\mathrm{p}K_a\leq7$ \cite{Car21}. Assuming the possible 
pH to be in the interval $1 \leq \mathrm{pH} \leq 12$, the corresponding 
adsorption energy parameter $\alpha$ would be within the interval 
$-25\lesssim\alpha=(\mathrm{p}K_a-\mathrm{pH})\ln 10\lesssim+25$. 
The value of the Flory lateral interaction strength $\chi$ is less certain 
and we are aware of only one instance where it was estimated from 
experimental data for ion induced lamellar-lamellar phase transition 
in charge regulated surfactant systems, where it can be on the order 
of a few 10 in dimensionless units of Eq.~\eqref{cr1} \cite{Har06}. 
These high values would be needed to overcome the electrostatic 
repulsion between similarly-charged lipids in this highly charged 
system. Without more detailed experimental input it thus seems 
reasonable to investigate the consequences of our theory for 
$-30 \leq \alpha \leq 30$ and $0 \leq \chi \leq 40$.

It is important to reiterate at this point that other charge regulation 
models are of course possible \cite{Pod18} and have been, apart 
from the protonation/deprotonation example, proposed for various 
dissociable groups in different contexts \cite{Bor01}. Our reasoning 
in choosing the particular Frumkin-Fowler-Guggenheim isotherm 
was guided by its simplicity in the way it takes into account the salient 
features of the dissociation process on the membrane surface, and 
the fact that the implied phenomenology has been analyzed before 
in the context of charged soft matter systems \cite{Har06}.

\subsection{\label{sec:total}The total free energy density, equilibrium charge configuration and flexoelectricity}

Combining now the electrostatic free energy and the charge regulation 
free energy, we are led to the explicit form of the total free energy of 
our model system as 
\begin{align}
{\beta {\cal F}_{tot}(n_1,n_2;\eta_1,\eta_2;R_1,R_2)} 
= {\beta{\cal F}_{ES}(n_1,n_2;\eta_1,\eta_2;R_1,R_2)} 
+{\beta{\cal F}_{CR}(n_{1};\eta_1; R_1)
+\beta {\cal F}_{CR}(n_{2};\eta_2; R_2)}.
\label{Ftot}
\end{align}
In order to find the equilibrium state of the system, the total free energy 
Eq.~\eqref{Ftot} then needs to be minimized with respect to the variables 
$\eta_{1,2}$. Introducing the dimensionless total free energy as
\begin{align}
\tilde{\cal F} = \frac{{\cal F}_{tot} \kappa_D \beta e^2}{\varepsilon_w \varepsilon_0} \equiv {\cal F}_{tot} \times (4 \pi \kappa_D \ell_B)
\end{align}
where $\kappa_D = \lambda_D^{-1}$ is the inverse Debye screening 
length and $\ell_B$ is the Bjerrum length, it can be derived that 
$\tilde{\cal F}$ depends only on the dimensionless electrostatic potential 
$u = \beta e \psi$, dimensional radial coordinate $x = \kappa_D r$ and 
dimensionless surface density 
\begin{align}
\tilde n_{1,2} = \frac{n_{1,2} ~\beta e^2}{\kappa_D \varepsilon_w \varepsilon_0} = n_{1,2} \times (4 \pi \lambda_D \ell_B),
\label{defni}
\end{align}
and thus
\begin{align}
\tilde{\cal F} = \tilde{\cal F}\left( x_1, x_2, \tilde n_1, \tilde n_2; \eta_1, \eta_2 \right).
\end{align}
Instead of the $x_{1,2} = \kappa_D R_{1,2}$ one can just as well 
introduce $h\equiv 1/(\kappa_D R)$ and $\kappa_D w$, assuming 
$R_1 = R, R_2 = R + w$, which is what we will do when plotting and 
analyzing the figures. The minimization of $\tilde{\cal F}$ with respect 
to $\eta_1, \eta_2$, corresponding to thermodynamic equilibrium, then 
yields the degrees of dissociation on both membrane surfaces as a 
function of parameters $x_1, x_2, \tilde n_1, \tilde n_2$, {\sl i.e.} the 
curvature and the surface density of the dissociable lipids.

As will become clear when we present the numerical results, for some 
values of these parameters the equilibrium can be characterized as a 
charge symmetry broken state, corresponding to $\eta_1 \neq \eta_2$. 
In that case the two surfaces have different charge or can even become 
oppositely charged.

The existence of charge symmetry broken states of a curved membrane 
has important consequences, among which flexoelectricity deserves 
special attention \cite{Pet99, Ahm15}. Flexoelectricity is a general 
mechano-electric phenomenon in liquid crystal physics but has important 
consequences specifically in the context of lipid membranes as argued by 
Petrov and collaborators \cite{Pet02, Pet06}. In the small deformation 
continuum limit regime, the induced flexoelectric polarization is proportional  
to the membrane curvature and a simple Langmuir isotherm based 
charge regulation model was invoked as a possible microscopic origin 
for the flexoelectric coefficient in the  seminal work of Derzhansky and 
coworkers \cite{Hri91}. The magnitude of the out-of-plane flexoelectric 
surface polarization density $p_S$ is defined as 
\begin{equation}
p_S \sim \vert \sigma_1 - \sigma_2\vert w,
\label{flexo1}
\end{equation}
being proportional to the difference in the surface charge densities. In 
the linear theory the surface polarization density, as well as the difference 
in the two surface charge densities, are proportional to the curvature of 
the membrane \cite{Pet99}, but - as we will see - this is not generally the 
case for charge regulated membranes which would then present a case 
of soft non-linear flexoelectricity \cite{Den14} or at least flexoelectricity 
with variable flexoelectric coefficient.

\subsection{\label{sec:CR}Charge regulation boundary condition}

The thermodynamic equilibrium is now obtained by minimizing the free 
energy Eq.~\eqref{Ftot}. We get two equations that correspond to charge 
regulation boundary conditions for $i=1,2$
\begin{align}
\frac{\partial {\cal F}_{ES}(\sigma_1, \sigma_2)}{\partial \sigma_i} \frac{\partial \sigma_i}{\partial \eta_i} +  \frac{\partial {\cal F}_{CR}(\eta_i)}{\partial \eta_i} = 0.
\label{BC1}
\end{align}
By taking into account Eqs.~\eqref{sigma12} and \eqref{BC}, as well as 
the form of the charge regulation free energy Eq.~\eqref{cr1}  we then 
derive the dissociation isotherm as
\begin{align}
\ln{\frac{\eta_i}{(1-\eta_i)}} = \alpha + \chi\eta_i - 2\beta e\psi_{i} = V({\eta_i}, \psi_i),
\label{BC2}
\end{align}
which can be solved for $\eta_i = \eta_i(\alpha, \chi, \psi_{i})$. From the 
above equation, combined with Eq.~\eqref{alphadef} it follows that on 
the mean-field level electrostatic interactions directly renormalize the 
dissociation equilibrium constant 
$\mathrm{p}K_a \ln{10} \longrightarrow \mathrm{p}K_a \ln{10} - 2\beta e\psi$. 
The above boundary condition corresponds to the Frumkin-Fowler-Guggenheim 
adsorption isotherm \cite{Pod18,Koo20} which is often invoked as a model 
in charge regulation problems \cite{Maj18,Maj19,Avn19,Avn20}. In terms 
of the surface charge density, Eq.~\eqref{sigma12}, the adsorption isotherm 
for protonation ($+$ charge, basic headgroups) and for deprotonation 
($-$ charge, acidic headgroups) can be obtained as
\begin{align}
\sigma_{i}(e) = e n_{i} \frac{e^{\pm V({\eta_i, \psi_i})}- 1}{\left(e^{\pm V({\eta_i. \psi_i})} + 1 \right)}, 
\end{align}
with $i = 1,2$ still standing for the inner and outer surface of the membrane, 
while $e$ is positive/negative for basic/acidic headgroups. Obviously the 
charged/uncharged state of the protonized/deprotonized lipid headgroups 
corresponds to a different sign of $\alpha$ and $\chi$.

The boundary condition derived above, together with the solution of either 
full PB equation or the linearized DH version for the electrostatic potential 
constitute the basic equations of our model the solutions of which we 
address next.

\section{\label{sec:results}Results}

For all the numerical evaluations presented herein, we have used typical 
system parameters such as the Bjerrum length $\ell_{B}=0.74 \, \mathrm{nm}$, 
the membrane thickness $w=4 ~{\rm nm}$, the dielectric constant of water 
$\varepsilon_w=80$ and that of the lipid as $\varepsilon_p=5$. For the 
dimensionless lipid density at the neutral plane, $\tilde n_0$, defined in 
Eq.~\eqref{defni},  we took a reasonable value $\tilde n_0 = 7.65$ which 
for an aqueous uni-univalent electrolyte  solution of $140\,\mathrm{mM}$ 
salt concentration (inverse Debye length $\kappa_D=1.215\,\mathrm {nm}^{-1}$, 
or equivalently,  screening length $\lambda_D=0.823\,\mathrm{nm}$), 
corresponds to $n_0=1\,\mathrm{nm}^{-2}$. With fixed $\tilde n_0$, because 
of the scaling relations, Eq.~\eqref{defni}, changing the screening length 
implies also changing the lipid density at the neutral plane, $n_0$.

In making sense of the numerical results we need to take due cognizance 
of the fact that our continuous electrostatic formulation remains valid only 
for radii of curvature much larger then the thickness of the membrane, 
$R \gg w$. Because all of the distances are scaled by the Debye length 
this furthermore implies that once the screening length is chosen, the 
numerical results are consistent only for $h \times (\kappa_D w) \ll 1$.

The two interaction parameters $\alpha$ and $\chi$ can be varied within 
relevant ranges, $-30 \leq \alpha \leq 30$ and $0 \leq \chi \leq 40$ as argued 
before.  In order to convert the dimensionless paramaters to physical ones, 
we note from Eq.~\eqref{alphadef1} that $(\mathrm{p}K_a-\mathrm{pH}) = \alpha/ \ln 10$ 
and $\chi$ is given in thermal units. In addition, we also present the differences 
between the full non-linear PB theory and the linearized DH theory expanded 
to the second order in the curvature of the membrane. This comparison is important 
since the linearized DH theory expanded to second order in curvature is often used 
to quantify the electrostatic and curvature effects in membranes \cite{Saf21}. Clearly 
both calculations exhibit similar qualitative features, but can be quantitatively quite 
different. We scan the parameter space in order to identify the important phenomena 
connected with charge regulation, which is our primary focus here, but do not 
specifically apply our model to any particular lipid membrane system.

From our previous works \cite{Maj18, Maj19} on two CR surfaces interacting across 
an electrolyte solution we know that depending upon the values of the parameters 
$\alpha$ and $\chi$ the Frumkin-Fowler-Guggenheim adsorption isotherm can lead 
to a charge symmetry breaking transition, corresponding to unequal equilibrium 
charge of two surfaces. This happens as a result of a competition among 
the three major interactions present in the system: the adsorption energy of ions 
or protons onto the charge-regulated surfaces, interaction of neighboring protonated 
sites on the surface and the electrostatic interaction between two charge-regulated 
surfaces. While the last one is at play only for a pair of surfaces, the former two are 
relevant even for a single CR surface and it can be shown that they lead to equally 
deep minima of the grand potential for charge densitites differing in magnitude as 
well as in sign for each CR surface in the absence of the other. A pair of interacting 
surfaces then automatically adopts an asymmetric charge distribution owing to an 
electrostatic-attraction-mediated reduction of the system energy. The charge 
asymmetry was found to be highest around the line $\chi = -2 \alpha$ and the 
charge symmetry broken region persisted also in the neighborhood of this line 
in the earlier works \cite{Maj18, Maj19}. The present case, describing 
a dielectric layer sandwiched between electrolyte layers, differs from the model 
of Ref.~\cite{Maj18}, pertinent to an electrolyte layer sandwiched between dielectric 
layers in the sense that the region between the two charged phospholipid leaflets 
is a dielectric, impermeable to electrolyte ions, while the electrolyte solution 
fills the rest of the space. The salient features of the charging behavior thus 
display a rather different behavior.

In fact while in our system we still find a symmetry breaking transition in the 
charging fractions, $\eta_{1,2}$, the symmetry broken charge region and the 
symmetric charge regions occupy switched places in the $(\alpha, \chi)$ "phase 
diagram" if compared to the case of Ref.~\cite{Maj18}; the line $\chi = -2 \alpha$ 
and its neighborhood thus corresponds to the symmetric charge region, while 
the rest of the "phase diagram" is symmetry broken. In addition, the extent of 
the symmetric region also depends on the curvature of the membrane, $h$, 
in such a way that the larger the curvature the larger is the extent of the charge 
symmetry broken region. This trend starts already at very small values of 
curvature. We now elucidate these statements in all the relevant detail.

We note that the parameters relevant for the plots are the dimensionless 
dissociation energy $\alpha$, the dimensionless lateral pair interaction energy 
of occupied neighboring sites $\chi$, and the dimensionless curvature $h$, see 
Table~\ref{Table:1} for definitions. As already stated, when converting from 
dimensionless to physical quantitites, once the screening length is chosen, 
one can only consider numerical results for $h \times (\kappa_D w) \ll 1$.

\begin{figure}[t]
\centering
\includegraphics[width=9cm]{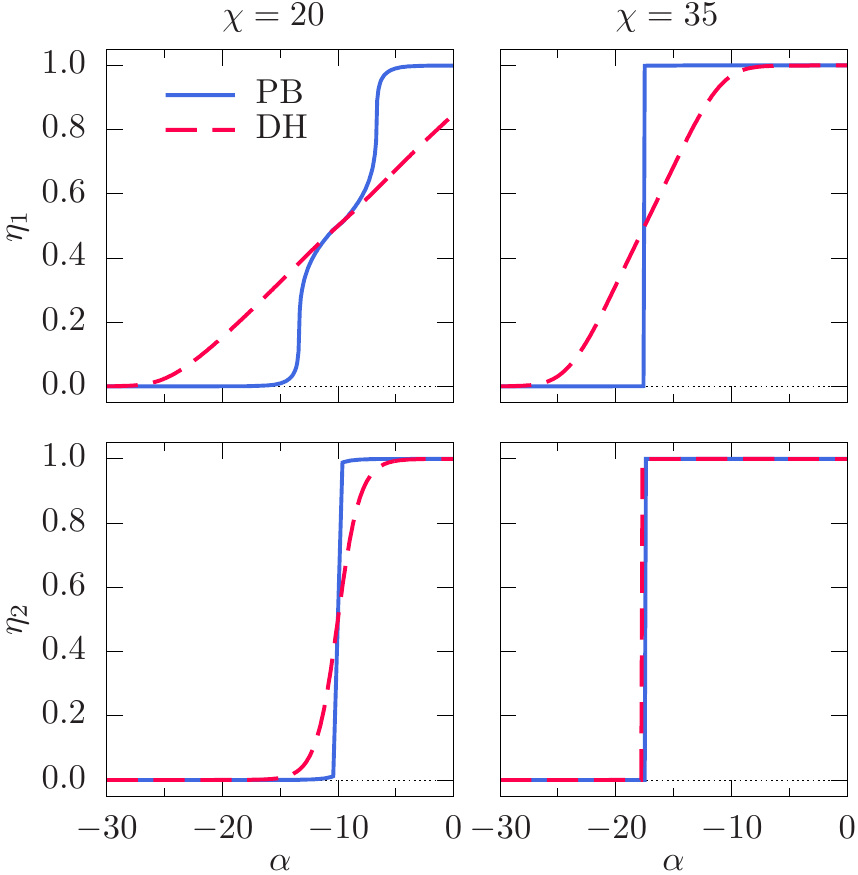}
\caption{\label{Fig:2}{The fraction $\eta_{1,2}$ of neutral lipid heads 
that are charged as function of the dimensionless adsorption/dissociation 
energy $\alpha$ at two different values of the strength $\chi$ of in-plane 
interaction between charged lipid heads for the dimensionless curvature $h=0.02$. 
Both $\alpha$ and $\chi$ are expressed in the units of the thermal energy $k_BT$. 
As one can infer from the plots, the discontinuous transition in at least on of the 
$\eta_{1,2}$ starts to show up at a smaller $\chi$-value within PB theory 
compared to that within the DH theory. In addition, the results also show 
symmetry-broken states ($\eta_{1} \neq \eta_{2}$) for $\chi=20$ within the 
PB theory and for both choices of the $\chi$-parameter within the DH theory.}}
\end{figure}

\begin{figure*}[!t]
\centering
  \includegraphics[width=13cm]{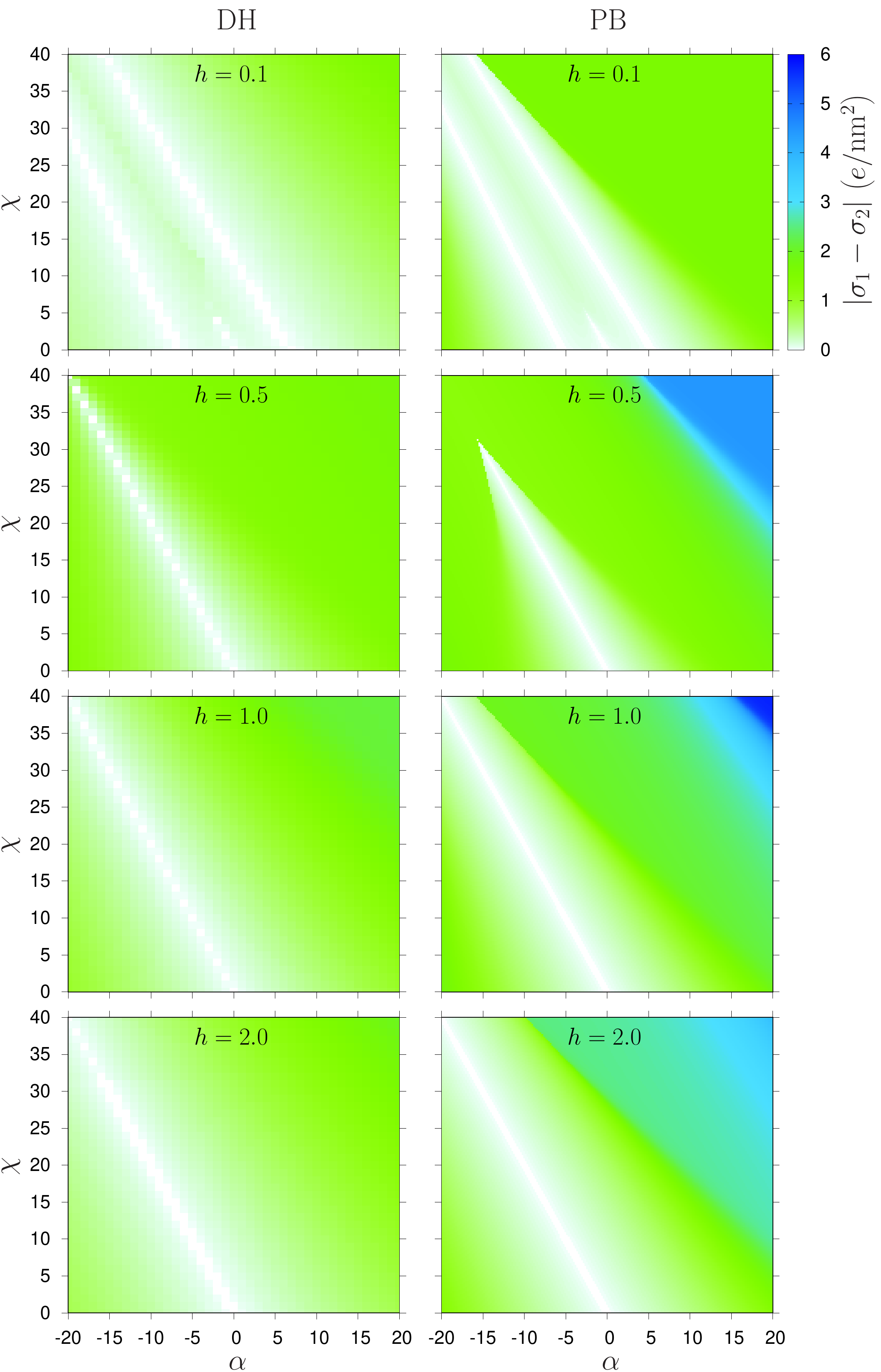}
  \caption{{The variation of the absolute value of the difference of the charge densities at the 
  two surfaces of the bilayer, $|\sigma_{1}-\sigma_{2}|$, expressed in the units of 
  $e/\mathrm{nm}^2$, as function of $\alpha$ and $\chi$ for four different choices of the 
  curvature $h = 0.1,\, 0.5,\, 1.0,\,2.0$ obtained within the DH theory (left panels) as well 
  as within the PB theory (right panels). The white regions in each plot correspond to 
  symmetric states with $\sigma_1=\sigma_2$ whereas the colored regions correspond 
  to asymmetric states with $\sigma_1\neq\sigma_2$.}}
  \label{Fig:3}
\end{figure*}

The two charge fractions $\eta_{1,2}$ as a function of $\alpha$ are displayed for different 
$\chi$ and a fixed membrane curvature $h$ in Fig.~\ref{Fig:2}. Clearly, for $\chi = 20$ even 
at the small curvature $h=0.02$ (corresponding to $R=\rm 41.15 ~nm$), both the PB and 
DH results show a pronounced charge asymmetry, which becomes even more prominent 
for higher values of $h$, see Fig.~\ref{Fig:3} for details. This charge asymmetry corresponds 
to the {\sl charge symmetry breaking} in the bilayer. In addition, the charge fractions are not 
only different but can exhibit a discontinuous transition as a function of $\alpha$. For 
$\chi = 20$ only the PB results show this transition for $\eta_2$, whereas the DH results 
do not; for $\chi = 35$ the PB results show a discontinuity for both $\eta_1$ and $\eta_2$, 
while the DH results show a discontinuity only for $\eta_2$ in the charge state of the lipids. 
As the curvature is increased the system remains in a charge symmetry broken state between 
the outer and inner leaflets of the membrane. We therefore conclude that for a spherical 
vesicle with finite, even if small, curvature the charge regulation standardly leads to a charge 
symmetry broken state. As a result the values for the charge fractions of the inner and the 
outer membrane layer differ even if the two leaflets of the lipid membrane are chemically 
identical. In addition, depending on the values of the charge regulation model parameters 
and curvature, the dependence of $\eta_{1,2}$ on either $\alpha$ or $\chi$ can be continuous 
or discontinuous. 

In order to be able to analyze the dependence of the charge densities $\sigma_{1}, \sigma_{2}$ 
of the two membrane leaflets on the charge regulation parameters, we present a phase diagram 
in Fig.~\ref{Fig:3} that shows the variation of $|\sigma_{1}-\sigma_{2}|$ as a function of 
$\alpha\in[-20,20]$ and $\chi\in[0,40]$ for $h=0.1,0.5,1.0,$ and $2.0$. The (white) regions, 
corresponding to charge symmetric states of the two leaflets, are clearly discerned and their 
location and extent depends on $\alpha$ and $\chi$, as well as on the the bilayer curvature 
$h$, which obviously has a profound effect on the charge state of the bilayer, {\sl i.e.}, on the 
value of $|\sigma_{1}-\sigma_{2}|$. We reiterate that in the previous work \cite{Maj18,Maj19} 
the symmetry broken region was centered on the line $\chi = -2 \alpha$, while in the present 
case it is the symmetric state which is centered on that line, while the rest of the phase diagram 
corresponds to a symmetry broken state. 

\begin{figure}[!b]
\centering
  \includegraphics[width=7cm]{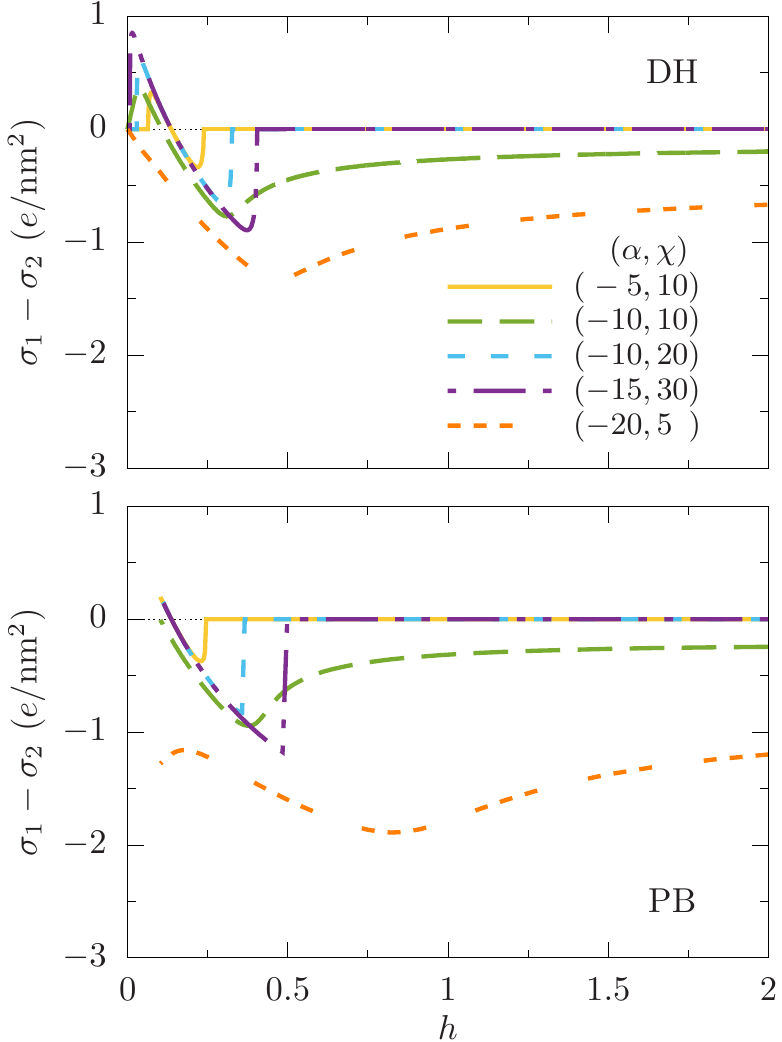}
  \caption{{Variations of the differences in the charge densities,    
           $\sigma_{1}-\sigma_{2}$, expressed in the units of $e/\mathrm{nm}^2$, as function 
           of the curvature $h$ for different combinations of the $\alpha$ and the $\chi$ parameters 
           with $\alpha$ being always negative (corresponding to a favorable adsorption free energy 
           for the protons onto the vesicle surfaces). Within both the theories (DH theory in the upper 
           panel and PB theory in the lower panel) the results are qualitatively the same, i.e., for 
           $\alpha$ and $\chi$ satisfying the relation $\chi = -2\alpha$, the charge density difference 
           $\sigma_{1}-\sigma_{2}$ shows a discontinuous transition from a finite non-zero value to 
           zero as the curvature $h$ is increased. However, for other combinations of the $\alpha$ 
           and $\chi$ parameters, the charge density difference $\sigma_{1}-\sigma_{2}$ do not 
           show any such discontinuous transition.}}
  \label{Fig:4}
\end{figure}

In addition, for $h=0.1$, we actually see not one but three regions of charge symmetric states, 
corresponding to $\sigma_{1}=\sigma_{2}$ (indicated by white color in Fig.~\ref{Fig:3}, almost 
coinciding for both non-linear PB and linear DH calculations. From the phase diagram it is difficult 
to see what the three symmetric branches correspond to, but we will later show that they 
correspond to the change in sign of the charge density difference, $\sigma_{1} - \sigma_{2}$. 
Among the regions with charge symmetry, the one centered on  $\chi=-2\alpha$ (the middle 
region) passes through almost the same range of $\alpha$ for both calculations. At $h=0.5$, 
the line $\chi=-2\alpha$ fully passes through the range $\alpha\in [-20,0]$ in the DH theory, 
while in the PB theory, the line is terminated at $\alpha\approx-16$. Further increasing the 
dimensionless curvature to $1.0$ and beyond, the line is terminated at $\alpha=-20$ for both 
theories. The more pronounced asymmetric states in Fig.~\ref{Fig:3} of the PB case extend 
over a broader region than for the DH case at $h=0.1$ and $0.5$. With increasing $h$ up 
to $1.0$ and higher, the phase diagram shows clearly that only the PB case can exhibit the 
highest asymmetry (represented by dark blue color, while the DH solution remains broadly 
less asymmetric. Also the PB theory exhibits a wider range of $|\sigma_{1}-\sigma_{2}|$ 
values than the DH solution.  

\begin{figure}[!b]
\centering
  \includegraphics[width=7cm]{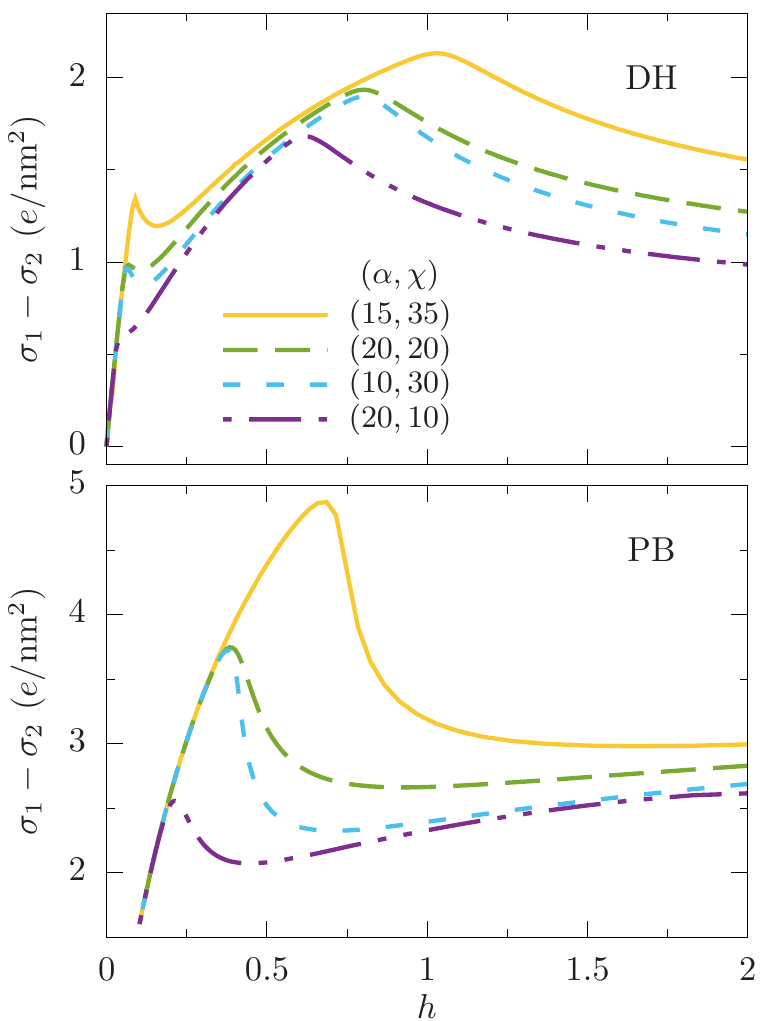}
  \caption{{Variations of the differences in the charge densities, $\sigma_{1}-\sigma_{2}$, 
  expressed in the units of $e/\mathrm{nm}^2$, as function of the curvature $h$ for different 
  combinations of the $\alpha$ and the $\chi$ parameters with $\alpha$ being always positive. 
  Within both the theories (DH theory in the upper panel and PB theory in the lower panel), one 
  observes a similar trend, i.e., the charge density difference $\sigma_{1}-\sigma_{2}$ increases 
  initially to reach a maximum, then goes down and increases again following a minimum. Within 
  the range of $h$ considered here, the results obtained within the DH theory shows the 
  occurrence of a second maximum as well.}}
  \label{Fig:5}
\end{figure}

We now turn our attention to the details of the curvature dependence of the charge state 
of the membrane. At the beginning a {\sl caveat} is in order: smaller values of the 
curvature are only accessible in the DH approximation, whereas the solution of the full 
PB theory take unreasonably long time because the system size has to be increases 
concurrently with the decrease in curvature. It is for this reason that the curvature 
dependence in the PB theory is truncated at finite values of curvature. 

The curvature dependence of the difference between the inner and outer charge density, 
$\sigma_{1}-\sigma_{2}$, for negative and positive values of $\alpha$ is fully displayed 
in Figs.~\ref{Fig:4} and \ref{Fig:5}, respectively. Obviously this dependence is fundamentally  
non-monotonic, including the changes of sign. For negative values of $\alpha$, see 
Fig.~\ref{Fig:4}, the difference $\sigma_{1}-\sigma_{2}$ is a non-monotonic 
function of the curvature exhibiting regions of unbroken charge symmetry, corresponding 
to $\sigma_{1}=\sigma_{2}$, as well as regions of broken charge symmetry with 
$\sigma_{1} \neq \sigma_{2}$, with the difference between the two surface charge densities 
varying from positive values to negative values on increasing the curvature. In fact this 
behavior can be seen clearly also in Fig.~\ref{Fig:3} where it corresponds to a curvature 
cut through the three lines of charge symmetry for a fixed $(\alpha,\chi)$ combination in 
the phase diagram. {Clearly for a sufficiently negative $\alpha$, {\sl e.g.} for $(\alpha,\chi)$ 
combination $(-20, 5)$, the difference between the inner and outer charge density, 
$\sigma_{1}-\sigma_{2}$ ceases to change sign within both the theories, but their 
non-monotonic behavior is still retained.} The two thus represent separate features 
of the charging state of the curved bilayer. The behavior for the positive values of 
$\alpha$, see Fig.~\ref{Fig:5}, differs significantly. For a variety of  $(\alpha,\chi )$ 
cuts through the phase diagram we detect here no changes in sign for the difference 
between the inner and outer charge density, but we do see remarkable non-monotonicity 
in its dependence on curvature with very clear-cut differences between the PB and DH 
results. Nevertheless, $\sigma_{1}-\sigma_{2}$ seems to increase for small curvature, 
reaching a local maximum, then dropping and increasing again but less steeply for larger 
curvatures. As for the range of curvatures that we display on Fig.~\ref{Fig:5}, one also 
needs to consider the inherent limitations of the continuum assumptions which form the 
basis of the present calculations and the curvature should not be extended to arbitrary 
large values.

The curvature dependence of the difference in surface charge density between 
the inner and outer leaflet, $\sigma_{1}-\sigma_{2}$, implies the existence of a 
{\sl dipolar moment} in the direction of the normal of the bilayer, see Eq.~\eqref{flexo1}, 
and therefore also the {\sl existence of flexoelectricity},  but with a variable magnitude 
and sign of the flexoelectric coefficient. Clearly, the region of a simple proportionality 
between the bilayer polarization and curvature is limited and depends crucially on the 
charge regulation parameters. 

For $\alpha > 0$, Fig.~\ref{Fig:5}, the difference $\sigma_{1}-\sigma_{2}$ does not in 
general change sign, but remains nevertheless non-monotonic so that the system 
remains in a broken charge symmetry state for all indicated values of the charge 
regulation parameters $\alpha$ and $\chi$. The calculations seem to indicate two 
separate regions of approximately linear flexoelectricity, but with  flexoelectric 
coefficients of different magnitude: one at small curvatures (below $h \simeq 0.5$ 
for PB calculations and $h \simeq 0.1$ for DH calculations) and another one at 
larger curvatures (above $h \simeq 0.75$ for PB calculations and $h \simeq 0.2$ 
for DH calculations), until finally $\sigma_{1} - \sigma_{2}$ varies only weakly with 
curvature. 

For $\alpha \ll 0$, Fig.~\ref{Fig:4}, the situation seems more complicated and also 
the differences between the PB and DH calculations more evident. There appears 
a linear flexoelectric regime at very small curvatures that changes sign for larger 
curvatures (evident for {\sl e.g.}  $\alpha,\chi = -20,5$ or $-10,10$ case for PB and 
DH calculations, Fig.~\ref{Fig:4}). As $\chi$ is increased the linear flexoelectric 
regime for small curvatures is eventually cut short for large enough curvatures and 
the system fully restores its charge symmetry with vanishing flexoelectricity (evident 
for {\sl e.g.} $\alpha,\chi = -15,30$ case for PB and DH calculations, Fig.~\ref{Fig:4}). 
Since the PB calculation cannot probe the regime of very small curvature because 
of numerical problems we can rely only on the DH results that show a vanishing 
flexoelectricity also for very small curvatures at not too large $\alpha < 0$ and $\chi > 0$. 
The non-linear features of flexoelectricity just described are pertinent to the charge 
regulation model which takes into account certain salient features of the phospholipid 
protonation/deprotonation or other charge dissociation reactions at the membrane-electrolyte 
solution boundary. They do not appear in fixed charge models of membrane electrostatics. 
Furthermore it is clear that the property of flexoelectricity depends crucially on the 
membrane dissociation properties as well as the solution conditions, and is thus far 
from being a universal property of the membrane composition.

\begin{figure}[!t]
\centering
\includegraphics[width=7cm]{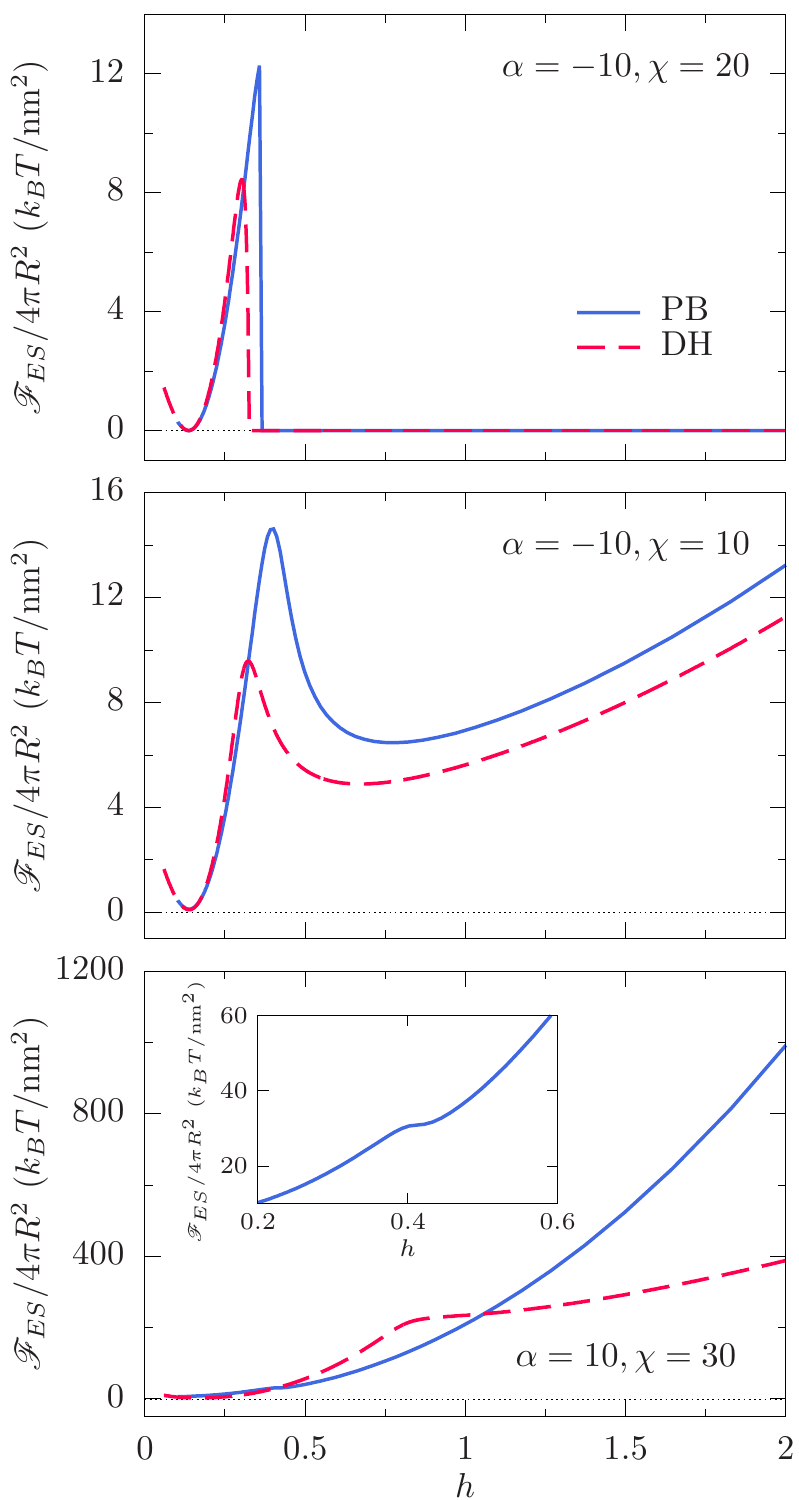}
  \caption{{Variations of the electrostatic part $\mathcal{F}_{ES}$ of the free energy 
  per unit inner surface area $4\pi R^2$, expressed in the units of $k_BT/\mathrm{nm}^2$, 
  as function of the curvature $h$ for different $\alpha$ and $\chi$ values. For a given 
  $(\alpha,\chi)$ parameter combination, the results obtained within the linearized PB 
  or DH theory are plotted using red dashed lines whereas those obtained using the 
  nonlinear PB theory are plotted using solid blue lines. Contrary to the widely used 
  approximation, $\mathcal{F}_{ES}/4\pi R^2$ does not in general show a 
  $h^2$-dependence. While it does mostly vary $\propto h^2$ or ${\propto (h-h_0)^2}$ 
  for $\alpha >0$, the situation is richer for $\alpha <0$ involving the presence of 
  multiple minima and even vanishing contribution corresponding to charge-neutral 
  surfaces for sufficiently large curvature values.}}
  \label{Fig:6}
\end{figure}

We finally examine the hypothesis that the electrostatic effects in membrane can be reduced 
to a renormalized spontaneous curvature and renormalized bending rigidity, standardly invoked 
in the context of electrostatic interactions in phospholipid membranes  \cite{And95,Dim19,Sho16}. 
In Fig.~\ref{Fig:6} we plot the curvature dependence of the equilibrium electrostatic surface free 
energy density as obtained from the linearized DH or fully non-linear PB theory. If indeed the 
electrostatic effects could be reduced solely to renormalized values of the bending rigidity and 
spontaneous curvature, then the curvature dependence of the equilibrium electrostatic surface 
free energy density should show a parabolic dependence centered at the spontaneous curvature. 
What we observe in Fig.~\ref{Fig:6} corroborates this expectations but only for positive values of 
$\alpha$. In general, and in particular for negative values of $\alpha$, however, the curvature 
dependence exhibits a behavior quite different from these expectations. In this case one 
observes, see Fig.~\ref{Fig:6}, multiple curvature minima and in some cases a vanishing 
value of the electrostatic free energy as both surfaces become completely neutralized. 
The paradigm of renormalized spontaneous curvature and renormalized bending rigidity 
has thus only a limited validity and its range of applicability depends again crucially on 
the phospholipid dissociation properties and solution conditions. The membrane composition, 
affecting the two charge regulation parameters, indeed plays a role in the curvature properties 
but so do the bathing solution conditions that collectively determine the nature and magnitude 
of electrostatic effects.

\section{\label{sec:discussion}Discussion}

There are of course a plethora of works based on molecular and coarse grained simulations 
of lipid bilayers and their interactions \cite{Fel00, Wan10, Bru16, Jos21} but none of these studies 
exhaustively explored the contribution of electrostatic interactions to the charge states of 
dissociable lipid headgroups \cite{Sch21}. Even in the most accurate parametrization of the 
lipid molecular force fields that take into account the dissociation state of the lipid headgroup 
through the partial charges, the charge regulation, {\sl i.e.}, the dependence of the dissociation 
state on the local bathing solution conditions, is usually not considered \cite{Yu21}. To include 
the effects discussed in our paper one would need to implement the charge dissociation 
reactions explicitly into the simulation scheme along the lines of recent advances in the 
simulation techniques for reaction ensembles \cite{Lan20, Cur22}. Inclusion of the explicit 
charge regulation model into the detailed mean-field electrostatic theory applied to a curved 
phospholipid bilayer membrane allowed us to uncover and analyse several effects that have 
heretofore remained outside the paradigm of electrostatic renormalization of the membrane 
mechanical properties \cite{Dim19}. 

The first effect depending on the detailed charge regulation mechanism is the charge symmetry 
breaking, leading to unequal charges of the inner and outer phospholipid membrane surface 
even if - and this is important - they be chemically identical, {\sl i.e.} described by the same free 
energy parameters. This effect was first observed in the case of membranes interacting across 
an electrolyte solution \cite{Maj18, Maj19}, however, the charge symmetry breaking for a curved 
bilayer differs from this case since the two charged surfaces in a membrane interact across 
a simple dielectric, and charge symmetry breaking is in some sense inverse to the case of 
interacting membranes. In the most drastic case the charge symmetry broken state can be 
characterized by one of the monolayers near neutral and only one charged (see the case 
corresponding to $\chi=20$ and $\alpha\approx -9$ in Fig.~\ref{Fig:2}, for example). 

The second important effect is the existence of  non-linear flexoelectricity \cite{Den14}, with 
flexoelectricity itself being well known and even explained by a type of charge regulation 
mechanism for membranes \cite{Hri91}. While sophisticated electrostatic models have 
been formulated for flexoelectricity analysis \cite{Lou13}, more detailed charge regulation 
description and full mean-field electrostatic theory indicate that the flexoelectric constitutive 
relation is in general not linear and in addition its form depends crucially on the charge 
regulation parameters. Only certain intervals of dissociation parameters would correspond 
coarsely to a linear flexoelectric constitutive relation, with dipolar moment proportional to 
the curvature \cite{Lou13}. In general, however, the proportionality is non-linear or there 
might actually be no flexoelectricity for sufficiently large curvatures.

The last important modification in our analysis of the charge regulation effects is the dependence 
of the free energy on the curvature. As stated before, the prevailing paradigm is to see the 
electrostatic effects as commonly renormalizing the elasto-mechanical properties of the 
membrane, such as surface tension, curvature and bending rigidity (for a recent description 
see Ref.~\cite{Dim19}). Our results indicate that this is not the complete story and that the 
free energy of a charge-regulated phospholipid membrane exhibits a much richer variety of 
curvature dependence. Only for a limited interval of the phospholipid dissociation parameters 
does one indeed observe a clear quadratic dependence on curvature that would be consistent 
with the electrostatic renormalization of the membrane elasto-mechanics. 

The bilayer membrane asymmetry need not necessarily involve the obviously asymmetric 
protein distribution in biological membranes. It can also exhibit differences in lipid composition 
between the two leaflets \cite{Ing14} or be a consequence of different solution conditions across 
the separating membrane \cite{Kar18}. Based on our calculations we can state that even without 
{\sl any} compositional asymmetry, or {\sl any} solution asymmetry, the phospholipid dissociation 
coupled to electrostatic interactions and curvature would itself contribute an essential charge 
asymmetry to the otherwise completely symmetric membrane properties. 

Experimentally, the very same model that we used above to describe the charge regulated membrane 
electrostatics was used to elucidate the liquid-liquid ($L_{\alpha} \longrightarrow L_{\alpha'}$) 
phase transition observed in osmotic pressure measurements of certain charged amphiphilic 
membranes \cite{Har06}. The phenomenon of charge symmetry breaking, that can induce 
attractions between chemically identical membranes, was argued to induce an attractive 
disjoining pressure in plant thylakoid membranes and photosynthetic membranes of a family 
of cyanobacteria \cite{Maj19}. We are thus confident that the charge regulation effects coupled 
to membrane curvature are real and could be detected in solutions of varying acidity and salt 
activity. While there is no lack in experiments probing the effects of pH on membrane properties 
of giant unilamellar vesicle such as a pH change induced vesicle migration and global 
deformation \cite{Kod16}, vesicle polarization coupled to phase-separated membrane 
domains \cite{Sta12} and the effect of localized pH heterogeneities on membrane 
deformations \cite{Kha11}, a systematic quantitative comparison between expected 
pH-induced effects and observed membrane response is lacking. One of the reasons 
for this is that the pH-curvature coupling is complicated and non-linear, as we argued 
and demonstrated above. The phenomenology of this coupling presented in this work 
should help to elucidate the details of the pH effects and the role played by the dissociation 
mechanism in phospholipid membranes.

While the measured effective rigidity of lipid membranes seem to corroborate 
the paradigm of  electrostatic  renormalization  of  elasto-mechanical  properties of 
membranes \cite{Dim19}, the experimental work of Angelova and collaborators, quantifying 
the effects of pH changes on lipid membrane deformations, polarization and migration of lipid 
bilayer assemblies seem to present a more complicated picture \cite{Sta12,Kod16,Ang18}. 
They demonstrated that a global or a local pH change can induce localized deformations 
of the membrane as well as the whole vesicle, polarization in membranes with phase 
separated lipid domains as well as whole vesicle migration. The models of pH effects 
used in this context are usually not based on explicit free energy contributions of charge 
dissociation, being the starting point of our work, but rather build upon assumed effective 
values of the membrane charges, and while the experimental studies do show the existence 
of different instabilities the detailed quantitative connection with our work is at present 
difficult to establish. Nevertheless, our elucidation of the coupling between the charging 
processes in lipid membranes and electrostatic interactions in general provide a solid 
underpinning for these type of phenomena which are clearly beyond the paradigm 
of electrostatic  renormalization of elasto-mechanical  properties of membranes.

Finally, our methodology and the model used have understandably and 
unavoidably many limitations. First of all, the limitations inherent in the PB continuum 
electrostatics, applying also to our theory, are well known and understood \cite{Mar21}. 
The assumption of incompressibily is relatively standard in membrane physics but of 
course entails certain well recognized limitations \cite{Dim19}. We only consider a 
quenched membrane state with fixed density and type of lipids, thus disregarding 
the annealing of the composition either by in plane diffusion or by trans-membrane 
flip-flops of different lipid species both associated presumably with large(er) time 
scales \cite{Hos20}. We do not exactly specify the composition of the membrane, 
neither its lipid part nor, possibly even more important, the membrane protein 
counterpart \cite{Jav21}, aiming for salient  characteristics of the behavior and 
disregarding the certainly important consequences of molecular identity \cite{Nap14}. 
We believe that irrespective of all these acknowledged limitations we uncover 
some features of membrane electrostatics which can be crucial in assessing 
and controlling the behavior of charged, decorated lipid vesicles. 

\section{Acknowledgments}

RP and PK acknowledge funding from the Key Project No. 12034019 
of the National Natural Science Foundation of China and the 1000-Talents 
Program of the Chinese Foreign Experts Bureau and the School of Physics, 
University of Chinese Academy of Sciences.  

%-------------------------------------------------------------------------------

\bibliography{Paper}

\end{document}